\documentclass[12pt]{article}     
  \usepackage{graphics}
  \usepackage{flafter}

\begin{document}
\title{Universal Velocity and Universal Force}
\author{Naresh Dadhich \footnote{email: nkd@iucaa.ernet.in}\\
IUCAA, Post Bag 4, Pune ~ 411~007, India}
\maketitle 
\begin{abstract}                      
In his monumental discoveries, the driving force for Einstein was, I believe, consistency 
of concept and principle rather than conflict with experiment. Following this Einsteinian dictum, we would first argue that homogeneity (universal character) of space and time characterizes 'no force' (absence of force) and leads to  existence of a universal velocity while inhomogeneity (again
 a universal property) characterizes curved spacetime and presence of a universal force which is present everywhere and always. The former gives rise to Special Relativity while the latter to General Relativity. 
\end{abstract} 

\section{Introduction}

Let me begin by hypothesizing that it is tempting to place Einstein or oneself in the 
pre-Maxwell times and follow the natural line of thought which is in the Einsteinian 
spirit, and see what happens. In this essay, I wish to follow the Einsteinian dictum of consistency of principle and concept, and let the rest follow naturally all by itself. 

We shall begin by probing the universal property of homogeneity of space and time, which would naturally lead to existence of a universal velocity and consequently to Special Relativity (SR). Also note that the universal character of space and time also leads to a universal constant velocity. As homogeneity charcterizes 'no force' (absence of all forces), inhomogeneity would characterize curved spacetime describing a universal force which is present everywhere and always. And the universal force is the Einstein gravity known as Generall Relativity (GR). 

\section{Universal Velocity}

We shall probe universality and homogeneity of space and time which would in a straightforward way lead to a universal constant velocity.

\subsection{Universality} The natural definition for  universal is that it is the same 
for all and equally shared by all. It could be an entity or concept like space, or could be a 
force or law like gravity. Since universal entity is the same and equally shared by all, this 
means all universal things must be related. That is, no two universal things can be 
independent. Any feature that distinguishes one universal thing from the other will 
have to violate the universal character. If there exist two universal entities or 
concepts, they must be related through a universal relation.  

What are the most primary universal entities we know of? The natural and obvious answer 
is space and time. They are indeed the same for all and equally shared by all. The two questions arise: since both are universal, hence one, they must be on the 
same footing and second, there must exist a universal relation between them.  

We know that the distance between two points in space depends upon the path an observer 
takes in going from one point to the other. It is a common experience that 
kilometer reading in a taxi is path dependent and that is why we are quite watchful 
that the driver takes the straight and not the circuitous path. Thus spatial distance is 
path dependent 
and so must be the time interval between any two events. This is what would be required 
to bring the two universal entities, space and time, on the same footing. This is however 
not so in the familiar Newtonian world. If we are to 
adhere steadfastly to our concept of universality, we are forced to seek new framework. 

Secondly, universality of space and time demands a relation between them which should 
also be universal. The natural relation between them is through velocity (which is 
formally defined by,  Velocity = Space/Time). A universal velocity is therefore required which is 
the same for all observers irrespective of their relative motion ~\cite{n0}. 

Thus universal character of space and time demands existence of a universal constant velocity which would bind the two into one, space-time. One of the natural consequences of the existence of universal velocity is that 
like spatial distance between two points, time interval between two events will now become path 
dependent. Universal velocity thus addresses both the above questions. However, like the path dependence of spatial distance which is a common experience, we do not observe the path dependence of time interval between events. For instance, take the two events as leaving home for work in the morning and returning in the afternoon. The clock at home and the clock one carries on one's wrist should in principle read different time between these two events because they have followed different paths. One has not moved in space at all  and remained stay put at home while the other has travelled in space to one's work place and back. Yet we see the two clocks reading the same. Could it be that the difference in time so small that it could be noticeable at eleventh decimal place which the clocks we use are unable to measure. Let us then do a thought experiment where the difference in two clocks becomes appreciable. 

Consider two observers A and B and let B go out on a space voyage in a spaceship travelling with a constant speed. He is told to go out for one hour according to his clock and then to turn back and come with the same constant speed. He is asked to send out light signal back to A every $6$ minutes. At what interval would these signals be received by A, would the interval be bigger, same or shorter? It would be bigger by a constant magnifying factor because B is moving away with a constant speed and hence light has to travel longer distance each time. Let the speed of B is such that magnifying factor is $3/2$;i.e. A will therefore receive the signals every $9$ minutes. The last signal A would receive from B's outward journey would be after $90$ minutes. Now B turns back with the same speed and hence the magnifying factor would be reversed to $2/3$ and so the signals would arrive to A now every $4$ minutes. When the two meet again, B's clock reads $60+60 = 120$ while A's reads $90+40 = 130$ minutes. I am often asked, which one is right? They are both right because like distance, time is also path dependent. 

[spacetime diagram: A triangle with two sides equal, verticle is t-axis and horizontal space axis. indicate 6 mts turning to 9 for outward and to 4 for inward journey] 

In the space-time diagram, the sum of two sides of the triangle is less than the third. This conflicts with our usual notion of geomtry. True, the space-time geometry cannot be the familiar Euclidean but instead it is hyperbolic which characterized by sum of two sides of a triangle being less than the third. The existence of universal constant velocity also requires a new geometry. 

\subsection{Homogeneity} Let us begin with Newton's First Law (NFL) which states that state of no motion (rest) is equivalent to state of uniform motion, constant speed in a straight line. The two are indistinguishable for any physical experiment or observation. This fact by implication means physics remains the same in all frames which are in uniform relative motion, the Principle of Relativity. Further any change from this situation can only occur by application or presence of a force. That is, NFL characterizes the force free state of motion and any departure from it is indicative of its presence. 

This is a universal statement which is indepedent of particle parameter and should therefore be true for everything. By universal we shall mean a property or statement which is true for all objects as well as for all reference frames (observers). This should also be true for zero mass particle which obviously cannot be at rest in any frame. It has to move relative to all observers and with a constant speed. Its speed should therefore be a limiting speed for all observers and hence a universal constant. If such a particle exists, it should move relative to all observers with the same constant speed in a straight line. If zero mass particle exists, it would have a universally constant velocity relative to all observers.

Another characterization of absence of force is that space is homogeneous and isotropic and time is homogeneous. Homogeneity of space means motion is completely free of the coordinates and hence they could be freely interchanged, for example, $x \to y$ and vice-versa. Since time is also homogeneous which means motion doesn't, like the space coordinates, depend upon time as well. As we could interchange $x$ and $y$, we should also be able to do the same for $x$ and $t$ for homeogeneity in space as well as in time. That is $x \to t$ but it couldn't be done  because their dimension does not match. If the concept of homogeneity has to be respected, this should happen and we have to match their dimension. That could only be done by asking for a universally constant velocity, $c$, so that $x \to ct$ and vice-versa. Once again we require a constant velocity which could be identified with the velocity of zero mass particle. Space and time are now bound together through this constant velocity and we now have the four dimensional space-time in place of space and time. NFL characterizes homogeneity of space-time which asks for a universal constant velocity ~\cite{n1}.

By homogeneity and universality of space and time, we are naturally driven to the existence of zero mass particles in nature propagating with a universally constant velocity. This is a profound prediction entirely dictated by concept and principle. If we admit them, we need a new mechanics because in Newtonian mechanics velocities add as $w=u+v$ which cannot keep any velocity constant for all observers. The new mechanics couldn't be anything different from Special Relativity (SR). SR could have in principle be discovered independent of and even before Maxwell's electrodynamics. If that were the case, it would have been the most remarkable discovery. That is, first a prediction based on general principles and then its actual verification provided by Maxwell's theory and  universal constant velocity being identified with the velocity of light. In its quantum description, light is the zero mass particle - photon. 

The universal statement that follows from NFL is that motion in absence of 
force (free motion) is always uniform in a straight line. This is a geometric statement and is therefore entirely the property of spacetime geometry without reference to any particle. It is given by geodesic (straight line) of $4$-dimensional homogeneous and isotropic spacetime described by Minkowski metric. This is how 'no force' is entirely synthesized in the geometry of spacetime.  

\section{Had I Been Born in 1844!} 

Had I been born in 1844, it could have been quite possible to follow this 
train of thought. Then in around 1870, I could have as a young man of 26 made the above profound prediction 
that there must exist a wave propagating in free space with a constant velocity. That 
would have been a good 5 years before the Maxwell's electrodynamics. It would have been 
amazing that you predict a wave in free space which Maxwell identifies with 
electromagnetic (light) wave and finally Hertz observes it experimentally ~\cite{n0}. 

Forget me, let us place Einstein as a young man in 1870, what he was 
in 1905 when he discovered SR, he could have very well argued as we did above and would have come up with the 
profound prediction before Maxwell's theory. Had that happened, it would have been a great display of 
pinnacle of human thought and Einstein as its purest human manifestation! 

However the events didn't take this chronology, yet it is insightful to wonder and probe the 
potential of sheer thought and analysis at its most pristine and sublime. This was precisely what was indeed displayed by Einstein while discovering GR - the most remarkable feat of human thought!  

\section{Universal Force} 

We postulate that there exists an interaction in which all particles interact with each other through a universal force. It is universal because of its universal linkage to all particles as well as its presence everywhere and always. It signifies a state dual (opposite) of no force which is never there while it can never be removed. It is always present and cannot be switched off or shielded from everywhere. As absence of force signifies homogeneity, similarly presence of force implies inhomogeneity. Since force is universal and hence it is always present for all particles, it implies that space-time has necessarily to be inhomogeneous (and/or anisotropic, in what follows we shall not state it explicitly, it would be taken as implied in the context). Therefore inhomogeneity should be incorporated in the structure of space-time in the same way as velocity of light is in the geometry of homogeneous space-time. Geometrically, what distinguishes homogeneous and inhomgeneous space is the Riemann curvature which vanishes for the former and it is termed flat while it is non-zero for the latter and it is then curved. Inhomogeneity thus produces curvature in space-time. Thus universal force curves space-time and hence it gets synthesized into space-time geometry. Its dynamics is now completely determined by the space-time geometry - its curvature.

Alternatively since it has universal linkage including zero mass particle (light). The massless particle, photon (light)  propagates  always with 
a universal constant velocity which cannot change. On the other hand, action 
of a force on anything is measured only through change in its velocity (motion). We have thus landed with a serious contradiction in principle. Since gravity is 
universal, it must act on massless photon as well, yet its velocity must not change. How 
is it possible? In the classical Newtonian framework, it is impossible to reconcile with 
these two opposing properties. What should we do? We have no other go except to expand the framework. How do we do that? 
   
When one is confronted with such a question of concept, it is only the robust common 
sense that can show the way. Let me take, an uneducated peasant as personification of 
uninhibited common sense. I ask him this question. After some ponder, he says that he 
can't much appreciate action of gravity on light, and asks back what would you want  
light to do in actuality to feel gravity? I reply that if light is grazing past a massive 
body, it should bend toward it as every other thing does. He breathes deep and hard,
scratches at his beard and then asks me to follow him, and takes me to the river which 
flows in the backyard of the village. He picks up a piece of log and throws into the 
river. It starts floating freely with the flow of the river. 

Then we walk along the river which suddenly bends, and so does the log. He smiles and 
inquires, did any force act on the log? No, I say, it is freely floating. But even then 
it bent with the river? He then triumphantly inquires, haven't you got the answer to your 
question? Dumb as I am, I say, No. He asks me where does your light float? In space, I 
say. Then what is the problem, he says, bend the space, light will bend automatically. Then it illuminates in me 
that why can't gravity bend/curve space around the massive body, and all particles 
including massless photons propagate/float freely in the curved space ~\cite{n2}. In fact curved 
space-time, as space and time have already been synthesized into one in SR.

Isn't this simply astounding? 

What have we arrived at? The universal force can honestly be described by 
no other means but by curvature of space-time itself. Then it  no longer remains a force 
but gets synthesized with structure of space-time and it simply becomes a property of the  
space-time geometry. Its dynamics, in particular the inverse square law,  
should now follow from the space-time curvature. That it does, all by itself, is what we show next.

Let us now turn to curvature of space-time and derive from that dynamics of the universal force. It is given by Riemann curvature tensor, $R_{abcd}$, which is constructed from the second derivative and square of the first derivative of the metric tensor, $g_{ab}$, defining the distance relation, $ds^2 = g_{ab}dx^adx^b$, for the $4$-dimensional space-time. It satisfies the Bianchi differential identity (analogue of curl of gradient, divergence of curl), the antisymmetric covariant derivative is identically zero, 
\begin{equation}
R_{ab[cd;e]} = 0  
\end{equation}  
where $;$ denotes covariant derivative and $[...]$ indicates antisymmetrization on the enclosed indices. If the equation of motion of the force has to follow from the curvature, it has to follow from this identity. The only thing we can do is to contract on the available indices which does lead, unlike for scalar and vector case, to a non-vacuous relation, 
\begin{equation}
G^{a}{}_{b;a} = 0, ~~~~ G_{ab} = R_{ab} - {1\over2}Rg_{ab} 
\end{equation}
where $R_{ab} = R^{c}{}_{acb}$ is the Ricci tensor, while $R = R^{a}_{a}$ is the trace of Ricci. Thus the trace (contraction) of the Bianchi identity yields a non-trivial differential identity from which we can make the following statement  
\begin{equation}
G_{ab} = \kappa T_{ab} - \Lambda g_{ab}, ~~~ T^{a}{}_{b;a} = 0 
\end{equation}  \\
where $T_{ab}$ is the second rank symmetric tensor with vanishing divergence,  the second term is a constant relative to covariant derivative, and $\kappa$ and $\Lambda$ are constants. The left hand side is a second order differential operator on the metric $g_{ab}$ (like $\nabla^2\phi$). For it to become an equation of motion, the tensor $T_{ab}$ should represent the source/charge for force. A source for universal force should also be universal;i.e. something which is shared by all particles and hence it should represent energy momentum distribution and the equation also ensures its conservation. With this identification, the above equation is Einstein's equation for gravitation. We have thus ended with Einstein gravity.

The universal interaction we postulated is in fact nothing but Einstein's gravity and its dynamics entirely follows from the space-time curvature. This is something very remarkable because no other force makes such a demand on space-time that it has to fully imbibe its dynamics. For all other forces, space-time provides a fixed inert background. It is the universality which integrates it into space-time structure. Since its dynamics is now property of space-time, we have no freedom to prescribe a force law, it all follows from the space-time curvature. In the weak field limit, the above equation reduces to the Newtonian equation, $\nabla^2\phi = 4 \pi G \rho$, signifying the inverse square law. It is important to note that Newton's inverse square law is contained in the Einstein equation and it is not prescribed but dictated by space-time curvature.

There are two constants in the equation of which $\kappa$ is to be determined by experimentally measuring the strength of the force and is identified with Newton's constant, $\kappa = -8 \pi G/c^2$. Why is there new constant $\Lambda$ which though arises in the equation as naturally as the energy momentum tensor, $T_{ab}$? It is perhaps because of the absence of fixed spacetime background which exists for the rest of physics and the new constant may be a signature of this fact. It should be noted that homogeneity and isotropy of space and homogeneity of time signifying force free state will in general be described by space-time of constant curvature and not necessarily of zero curvature. The new constant $\Lambda$ is the measure of the constant curvature of de Sitter (dS) or anti de Sitter (AdS) space. It may in some deep and fundamental sense be related to the basic structure of space-time. 

Gravity distinguishes itself from all other forces by the remarkable feature of 
impregnating space-time itself by its own dynamics. It then ceases to be an external force. 
Gravitational field is fully described by the curvature of space-time geometry. Motion under 
gravity will now be geodesic (straight line) motion relative to curved geometry of 
space-time. The geodetic motion should naturally include the Newtonian inverse square 
attractive pull, and in addition it should also have the effect of curvature of space. 
Light cannot feel the former but only feels the latter. Stronger is the gravity, stronger 
would be the curvature it produces. As we make the field stronger by making larger and 
larger mass confined to smaller and smaller radius, it is conceivable that space gets so 
curved that light cannot propagate out but its orbit closes on itself around the massive 
body. When that happens a black hole is defined from which nothing can propagate out. 
Things can only fall into but nothing can come out - it defines a one-way membrane, an event horizon. This 
is the most remarkable and distinguishing prediction of the Einsteinian gravity. 
  
We thus have both universal force and no force, which are dual to each-other, are fully incorporated in geometry of space-time, the former curved while the latter flat. This is very satisfying and what it indicates is the general principle that anything universal should ultimately be synthesized in space-time geometry.

\section{Universal Force and Newton's Second Law}  
We have obtained the equation of motion for Einstein's gravity from curvature of spacetime simply by following the differential geometric identity in a straightforward and natural manner. No reference was made to the celebrated Principle of Equivalence (PE) which served as a great motivation and played cornerstone role in Einstein's journey from SR to GR. It is based on mass proportionality of gravitational force, thereby the universality of acceleration by application of Newton's Second Law (NSL) for all massive particles. In the usual equation of motion, we have $m_i {\bf  \ddot x} = m_g \nabla \phi$ where $m_i$ is inertial mass and $m_g$ is gravitational mass. The fact that the acceleration could be anulled out in freely falling lift requires $m_i = m_g$. However, why should that be so and there is no conceivable physical reason for that? Finding the physical reason for the equality is in itself as formidable a problem as one can think of. It is simply being taken as an empirical observational fact without any explanation and understanding. On the other hand, it could be easily bypassed as we didn't have to make any reference to it in deriving Einstein's gravitational equation. It simply followed from space-time geometry. 

The question is application of NSL to gravity. Since universal force has also to link to massless particle, the equation of motion under gravity has therefore to be free of mass. The only way it can happen is when motion is given by geodesic of space-time geometry which incorporates force in its curvature. Once we have curved space-time, then its curvature itself dictates the dynamics of the force. Motion under gravity for both massive as well as massless particles is simply described by geodesics of curved space-time. The universality of acceleration of massive particles is because of geodetic motion in curved space-time and not because of equality of inertial and gravitational mass.$^{[1]}$\footnotetext[1]{The E$\ddot{o}$tv$\ddot{o}$s like experiments are interpreted to establish their equality to very high precision, one part in $10^{14}$. This could as well be interpreted as the experimental verification of gravity being described by curved space-time. However in GR, the gravitational field equation (3) in which energy monemtum (inertia), $T_{ab}$, is also the source for gravity, measure of gravitational charge and thereby in a sense indicating  $m_i = m_g$.} The question of their equlity therefore becomes irrelevant. The point to be noted is that NSL is not applicable for motion under Einstein's gravity. The curved space-time naturally incorporates PE in the property that at a given point it is always possible to define a tanget plane which is free of gravity giving local inertial frame (LIF). Since space-time is curved, there can exist no global inertial frame but only LIFs, and then Principle of Relativity says all LIFs are equivalent.

Thus PE becomes a property of curved space-time and not so much a driving force for Einstein's gravity. J. L. Synge was the first to voice this sentiment forcefully when he famously pronounced,

{\it " The Principle of Equivalence performed the essential office of midwife at the birth of general relativity, but, as Einstein remarked, the infant would have never got beyond its long-clothes had it not been for Minkowski's concept. I suggest that the midwife be now buried with appropriate honours and the facts of absolute space-time faced"} ~\cite{synge}. 

Let us reiterate that motion under gravity like no force is purely a property of space-time geometry and hence cannot be governed by Newton's laws of motion. This is an important fact not often emphasized upon. 

\section{Outlook}   

The main theme of the essay is to demonstrate the power and sheer simplicity of pure  
thought 
and robust common sense. It is a tantalizing hypothesis that had a young person in 1860s 
wondered about things and argued as we have done above, it would have been quite possible 
to predict existence of a wave propagating with universal constant velocity in vacuum, 
and consequently the new mechanics - Special Relativity. All this could have happened without any experiment challenging 
the existing theory. This was precisely what had happened for Einstein's discovery of 
General Relativity which was purely principle and thought driven. We are taking a one 
more step forward in that direction. Conceptually, all that what was required was available for SR not only in 1860s but right from Newton's time. 

Why do I then single out 1860s? Because before 1800s, the prediction would have been too 
much ahead of its time, and hence would have had a danger of being still born. If a scientific prediction is not experimentally verified in a reasonable timeframe of 30-40 years, it tends to be forgotten and it then generally belongs to the rediscoverer at an opportune time. Take the example of the 
greek philosopher, Aristochrus who is believed to have proposed that the earth goes 
around the sun and not the other way round as early as second centuray BC. $^[2]$\footnotetext[2]{In the Samos island of Greece, there does in fact exist his bust with the inscription that his discovery was later on copied by Copernicus in 1564.} This discovery couldn't 
survive for want of proper intellectual base and  understanding which came into existence 
only in 16th 
century through the observation of planetary orbits. Then the time was ripe for Copernicus 
to make the monumental discovery. It is in this context that 1860s and little earlier attain significance. 
With the Maxwell's theory of electromagnetism soon coming, the predicted wave 
would have been identified with the electromagnetic wave to be observed experimentally 
by Hertz. The stage was therefore well set for the profound prediction and discovery. 

It is interesting to conjecture and wonder, but the hard fact of life is that 
scientific discoveries are seldom driven, with the honorable and unique exception of 
General Relativity, by pure thought. They are essentially driven by contradiction 
between theory and observation, and the latter also depends upon the available technology 
for instruments. Truly, they are the products of the times - the prevailing scientific driving ideas, concepts and experimental backup. However had it happened as we envision, it would have been really 
great.

At any rate, it is a wonderful and enlightening way of looking at things 
based purely on simple logic and robust common sense. It is indeed insightful to see how much one can derive from the universal characteristics of space-time?

\end{document}